# On the origin and nature of double-double radio galaxies

David Garofalo, Zhiyuan Liu & Atticus V. Magerko

Department of Physics, Kennesaw State University, USA

Abstract

Double-double radio galaxies (DDRGs) display inner and outer jets or lobes thought to result from intermittent accretion. Due to randomly triggered accretion events, the lifetime of the retriggered jet is not expected to have any connection to the time of quiescence between jets, yet we show that a correlation between the two quantities may exist, which we interpret as resulting from continued accretion through the quiescent jet phase. Despite continuous accretion, a jet is absent because its presence depends on a non-zero value of black hole spin, but accretion transitions the system from counter-rotation to corotation, and therefore through zero black hole spin where a jet cannot form. The time of jet quiescence depends on how long it takes to spin the black hole up again in corotation, which is longer for lower accretion rates. Once the black hole spin is large enough for a renewed jet, this inner jet will last longer the lower the accretion rate is. Hence, in a continuous accretion scenario, longer quiescent times tend to associate to longer inner jet times. In addition, DDRG jets are of FRII morphology which we show to result from the absence of a tilt in the accretion disk in the transition through zero black hole spin, ensuring the absence of an FRI jet in a way that connects with our understanding of X-shaped radio galaxies. Both correlated timescales as well as sameness in jet morphology offers evidence in favor of our picture.

## 1. Introduction

One of the cornerstones of our current picture for active galaxies is the idea that black holes are triggered at random times by chaotic or secular processes in galaxies or by galaxy interactions (Schoenmakers et al 2000) which then provide a feedback effect (e.g. Fabian 2012) that explains the differences observed as a function of environment and whether a jet is produced (Moderski et al 1998; Wilson & Colbert 1995; Sikora, Stawarz & Lasota 2007). Double-double radio galaxies (DDRGs – see Figure 1) fit in this picture as active galaxies whose triggers happen to produce jets with lobes, the signatures of which last long enough for us to detect the occurrence of a sequence of two jet phases (Saikia & Jamrozy 2009; Kuzmicz et al 2017; Mahatma et al 2019). In this picture, the quiescent phase – i.e. the time during which the jets are off – is simply a period during which the black hole is not accreting. If this picture were correct, there would be no correlation between the lifetimes of either the inner or outer jet and the time of quiescence. Marecki et al 2021 report data on the lifetime of outer and inner jet lobe signatures and times of quiescence in a sample of 10 DDRGs allowing us to explore this question (see also Konar et al 2012). We find that a correlation exists between the time signatures of the inner, later jet, and the time of quiescence for the average values in the data. This points away from the random trigger as the explanation for DDRGs. Given the small sample size, we found two additional sources. While a correlation exists in the data, the statistics are not robust, and we do not wish to claim strong support for our picture. Instead, we see this as an opportunity to open up a conversation about the possibility of a physical explanation in terms of continuous accretion through the quiescent jet phase in the context of a model for black hole accretion and jet formation that posits retrograde accretion as a fundamental element in the generation of powerful radio galaxies. This is the first time we attempt to include DDRGs in this framework. In addition to this, our theoretical framework allows us to also understand the nature of jet morphology in DDRGs which we show to be connected to jet orientation. The FR jet morphology was initially divided into FRI and FRII (Fanaroff & Riley 1974). Over the last half decade, in fact, we have come to recognize that a compelling picture for understanding FRI jet morphology emerges because of the transition through zero black hole spin and the possibility that accretion may change direction as the corotating phase ensues (Garofalo, Joshi et al 2020; Singh et al 2024). If the accretion disk changes direction, our paradigm prescribes that an FRI jet will form and otherwise not (Singh & Garofalo 2023). In Section 2.1 we analyze the observational evidence and show the correlations. In Section 2.2 we provide the theoretical picture and the origin of the physics of the statistics behind the putative correlation. In Section 3 we describe DDRGs more generally by fitting them within our paradigm and conclude.

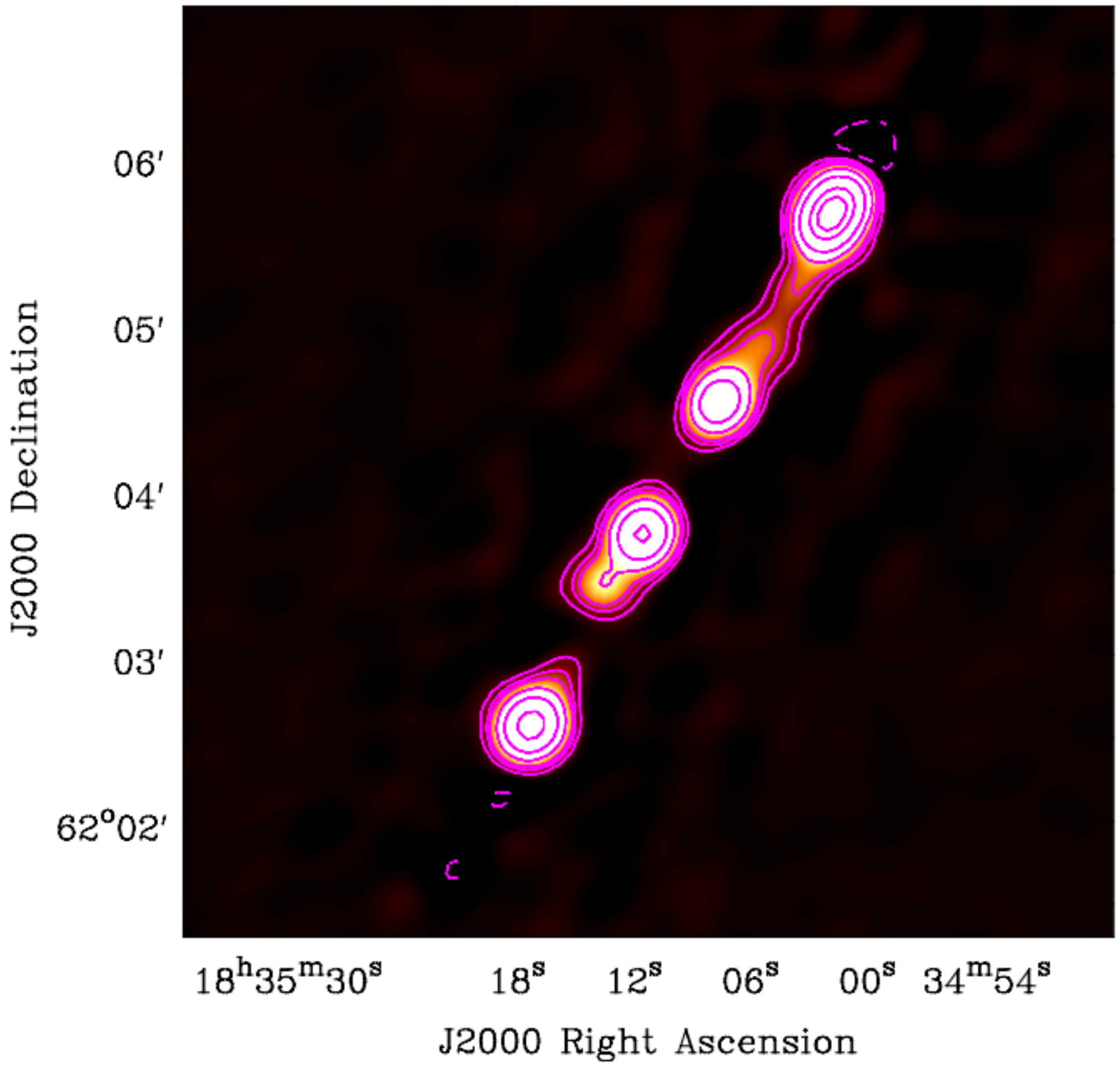

Figure 1: LOFAR 144 MHz image of the DDRG J1835+6204 from Orru' et al. 2015.

## 2. Double-double radio galaxies

### 2.1 Double-double radio galaxies - observations

The jet and quiescent lifetimes of ten DDRGs are reported in Marecki et al. (2021) Table 7 and reported here in Table 1 plus two additional sources, 3C 219 and 3C 388.

| source name | z | $t_{in}$ (Myrs.) | $t_{out}$ (Myrs.) | $t_{quies}$(Myrs.) |
|---|---|---|---|---|
| J0028+0035 | 0.3985 | 3.6 | 245 | 11 |
| J0041+3224 | 0.45 | 4.0 | 105 | 11 |
| J0116-4722 | 0.1461 | 1-28 | 66 – 236 | 1.4 – 65.4 |
| J0840+2949 | 0.0647 | 0.12 – 33 | > 200 | 2.0 – 102.0 |
| J1158+2621 | 0.1121 | 0.5 - 4.9 | 113 | 6.6 – 11.0 |
| J1352+3126 | 0.045 | 0.3 | 62 | 0.7 |
| J1453+3308 | 0.2482 | 5.0 | 104 | 24 |
| J1548-3216 | 0.1082 | 9.2 | 132 | 30 |
| J1706+4340 | 0.525 | 12 | 260 – 300 | 27 |
| J1835+6204 | 0.5194 | 1.34 – 2.25 | 22 | 1.0 – 6.6 |
| J0921+4538 (3C 219) | 0.1747 | 0.15 | N/A | 0.5 |
| J1844+455 (3C 388) | 0.091 | 6 | N/A | 4 |

Table 1: The 10 sources listed by their J2000 name, redshift, and the lifetimes for the inner and outer jet, and the time of quiescence from Marecki et al. 2021 plus two additional sources. 3C 388 is from Brienza et al. 2020; 3C 219 is from Wolnik et al. 2024. The references for the 10 objects are Machalski et al. (2011), Konar et al. (2013), Jamrozy et al. (2007), Machalski et al. (2016), Marecki et al. (2016), Konar et al. (2012), and Konar & Hardcastle (2013).

We choose to plot average values. In Figure 2 we plot $t_{in}$ versus $t_{quies}$ for the objects in Table 1. In Figure 3, we plot $t_{in}$ versus $t_{out}$ from Table 1 for the sources in Marecki et al. 2021, and in Figure 4 we plot $t_{quies}$ versus $t_{out}$. The Pearson coefficient is 0.926 and the low p-value of $10^{-5}$ for the data in Figure 2, while interesting in that it points to a connection between $t_{in}$ and $t_{quies}$ that should not exist in the context of a time random jet trigger for the inner jet, the statistical significance is not sufficiently high to make robust claims. The idea that the data may correlate this way is tantalizing and our goal is to begin a conversation that will encourage observers to look at many other DDRGs in this way. The absence of correlation in Figures 3 and 4 is captured by the low Pearson coefficients 0.587 and 0.514, respectively. In Figure 2 we show both vertical as well as the horizontal range in the data, which we now discuss.

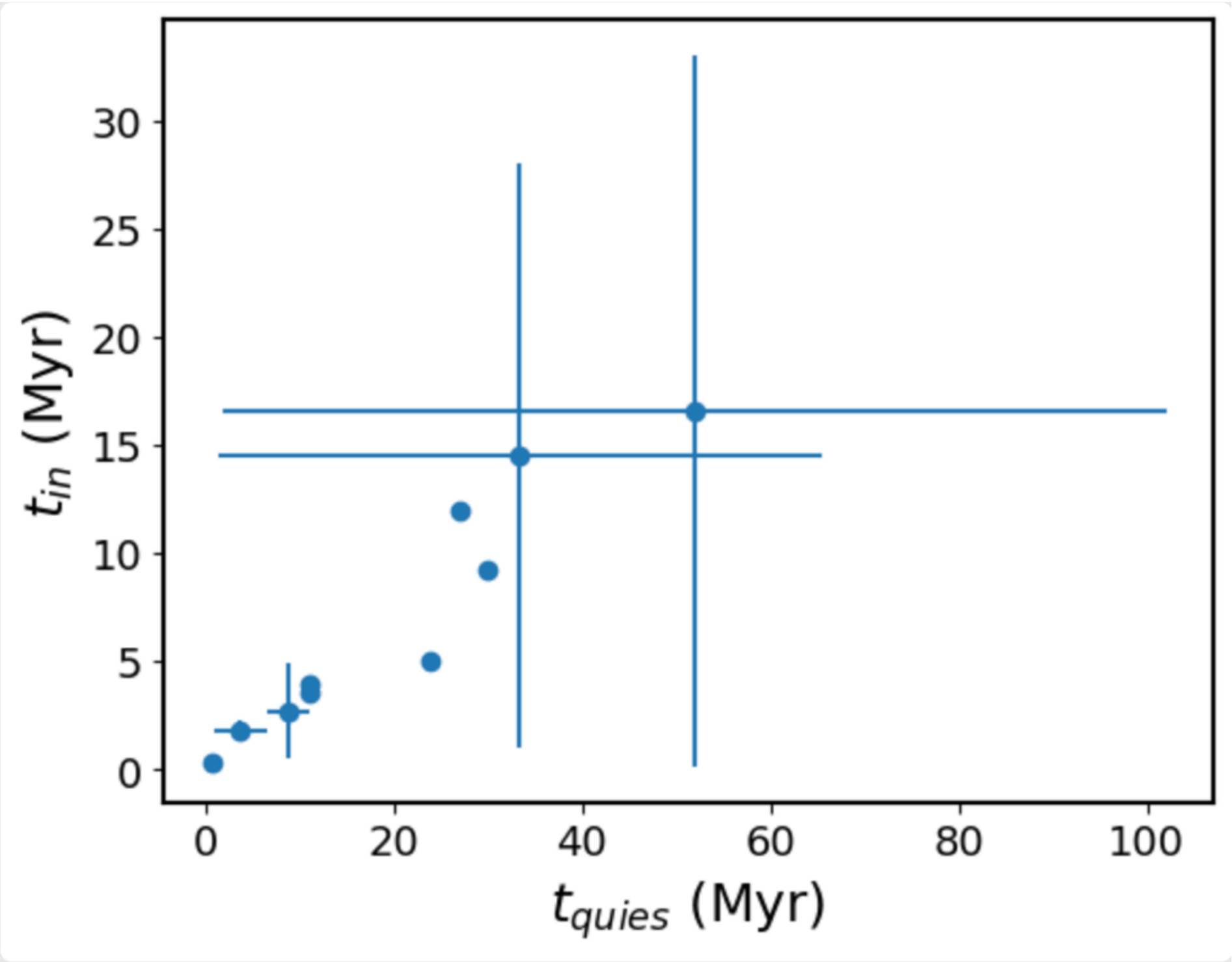

Figure 2: $t_{in}$ versus $t_{qui}$ with a Pearson coefficient of 0.926 and a p-value of 1.5 x $10^{-5}$. Range of values reported in the literature associated with each data point.

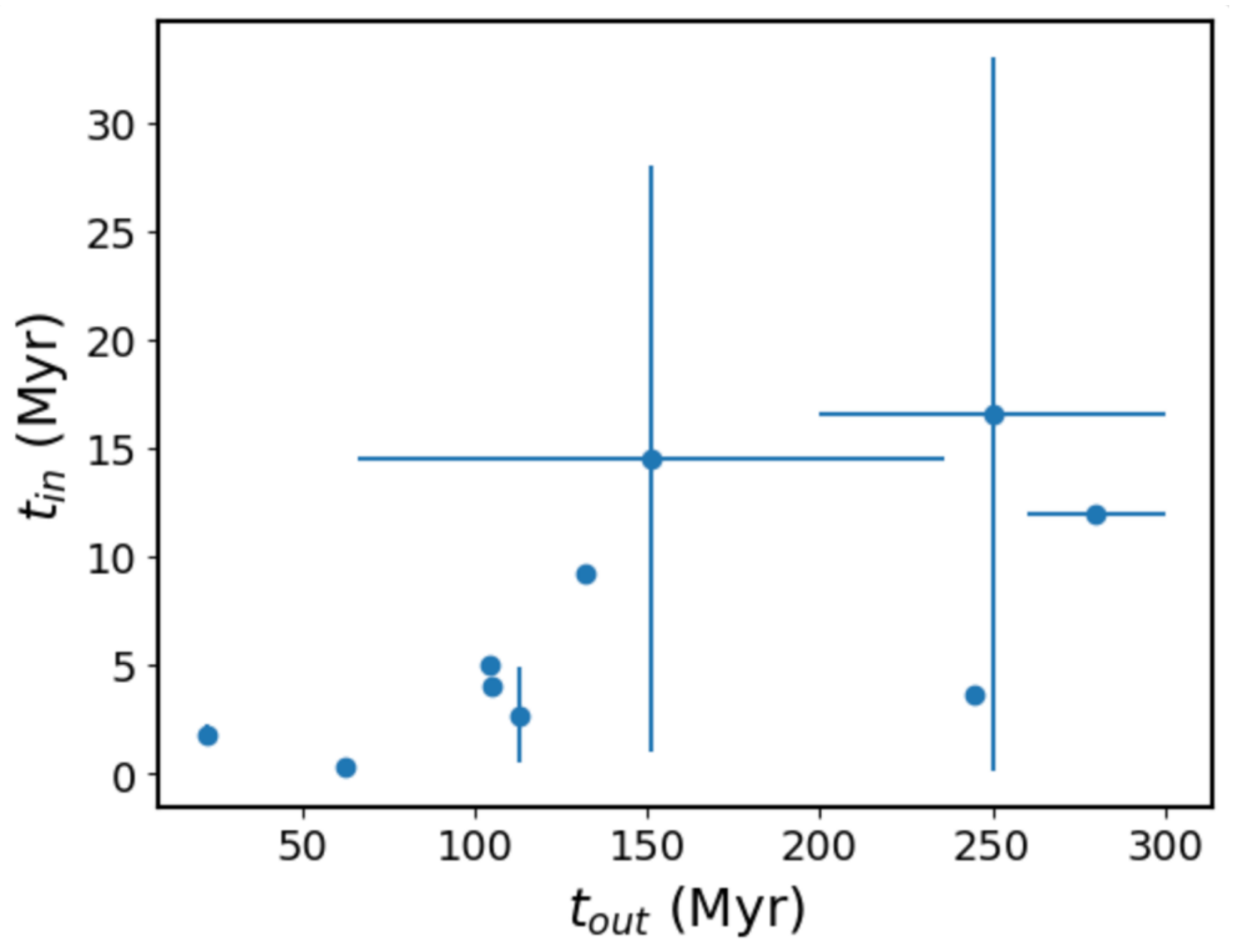

Figure 3: Absence of correlation between $t_{in}$ and $t_{out}$. Pearson coefficient equal to 0.547. Range of values reported in the literature associated with each data point.

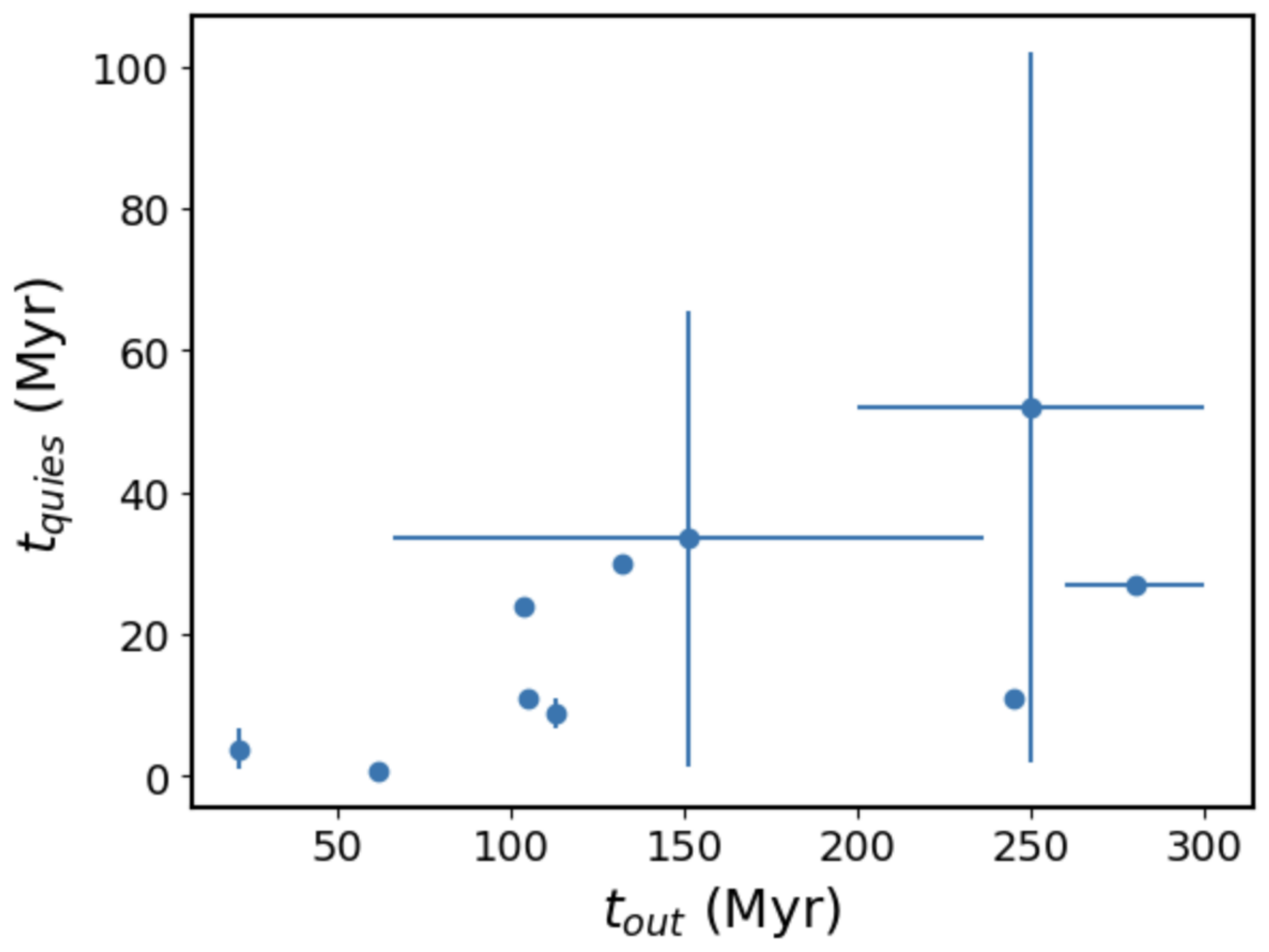

Figure 4: Absence of correlation between the quiescent time and the lifetime of the earlier jet. Range of values reported in the literature associated with each data point.

The sample compiled by Marecki et al. (2021) appears to compare the parameters they derived for the source they studied in their paper with other sources from the literature. The assumptions and methodology adopted by the different authors referred to in the references cited by Marecki et al. (2021) for the list of sources vary widely. Model parameters used by Machalski et al. (2011) referred to in the compilation differ from those of Brocksopp et al. (2011). For J0840+2949, Marecki et al. (2021) refer to the paper by Jamrozy et al. (2007) but while the period of quiescence is reported in Jamrozy et al. (2007) to be less than about 100 Myr., they report a range of values from a lower estimate of 2 Myr. all the way up to 102 Myr. This source has the highest value in the compilation. The values quoted in Marecki et al. (2021) for the quiescent time scale of J1835+6204 are 1.0–6.6 Myr. taken from Konar et al. (2012). These values estimated by Konar et al. were under the assumption that the North-Western hotspot, which has a prominent hotspot unlike the South-Eastern one, is on the far side. This is questionable as the location of the hotspots relative to the core do not appear to be consistent with this assumption. These caveats make it clear that the range of time estimates cannot be considered as uncertainties in any rigorous sense in Figures 2 through 4 but are the result of varying the assumptions. These issues plus the small sample size do not allow any robust conclusions to be drawn. While the correlation is interesting and the data tantalizing, its statistical significance is weak, and this study should therefore be viewed, again, more along the lines of an attempt to begin a conversation. Given the existence of a few hundred DDRGs and giant DDRGs, we hope that observers will investigate this further.

Unlike X-shaped radio galaxies whose later jet may be of FRI morphology, DDRG inner jets share the same morphology as their outer jet counterparts. They are both FRIIs (Schoenmakers et al. 2000; Konar & Hardcastle 2013). The issue is whether a DDRG could be formed with an inner jet of FRI morphology. One problem is overcoming selection bias. Detecting the hot spots of FRIIs as opposed to the diffuse emission in FRIs is easier and so there is the potential for mislabeling potential DDRGs with FRI inner jets with simple FRII radio galaxies, especially in light of the resolution of FIRST (e.g. Mahatma et al. 2019). A more global and deeper observational investigation of FRII radio galaxies would need to be undertaken to explore this issue. In the next section, we apply our paradigm explaining why there should be a correlation between $t_{in}$ and $t_{qui}$ and why both jets are expected to share the FRII morphology.

## 2.2 Double-double radio galaxies - theory

The basic theory behind all jetted AGN that we adopt here comes from Garofalo, Evans & Sambruna 2010. Since the most powerful FRII radio galaxies are triggered, according to this model, by accretion in counter-rotation around a highly spinning black hole, there is an intrinsic black hole spin-down followed by spin-up in corotation if accretion fuel is sufficient. As a result of this, there is a threshold black hole spin value in counter-rotation for which the jet "turns off" or is no longer visible, and a threshold black hole spin value in corotation for which a new jet "turns on" or becomes visible.

### 2.2a Timescale of jet quiescence

If we consider the jet to be "off" for spin values in the range $-0.1 < a < 0.1$ (where negative dimensionless spin $a$ represents counter-rotation and positive dimensionless spin $a$ represents corotation), the time to cross that spin range at the Eddington limit is at least a few million years (Garofalo et al. 2020). Since the model predicts that accretion will not violate the Eddington-limit but may drop below that over time depending on feedback, we can derive a theoretical limit for the quiescent time. To accomplish this, we need to find the rate at which the dimensionless black hole spin changes with time, $da/dt$. The dimensionless black hole spin is related to the angular momentum $L$ of the black hole via

$$a = cL/(GM_{BH}^2) \qquad (1)$$

with G Newton's constant, c the speed of light and $M_{BH}$ the black hole mass. We write this in terms of the infinitesimal increment in angular momentum delivered to the black hole by accreted gas as

$$dL = GM_{BH}^2 da/c. \qquad (2)$$

A parcel of gas dm that is moving on circular orbits with speed v that reaches the disk inner edge at r, carries an infinitesimal amount of angular momentum. Thus, in magnitude

$$dL = dm\ v\ r. \qquad (3)$$

Our goal is to reframe this in terms of the dimensionless black hole spin. This gives

$$GM_{BH}^2 da/c = dm\ v\ r\ . \qquad (4)$$

On circular orbits and using Newtonian theory, we have

$$v = (GM_{BH}/r)^{0.5}. \qquad (5)$$

The value of r at the inner edge varies as a function of black hole spin but for simplicity we consider it to be constant, especially because it varies from slightly larger than $r = 6GM_{BH}/c^2$ for $a = -0.1$ to slightly below that value for $a = 0.1$. In other words, we use the value of r associated with zero black hole spin as the average inner disk radial location. This gives

$$GM_{BH}^2 da/c = dm\ (GM_{BH}/r)^{0.5} 6GM_{BH}/c^2 = dm(6)^{0.5}\ GM_{BH}/c \qquad (6)$$

which then leads to,

$$da/dt = M_{BH}^{-1}(6)^{0.5}\, dm/dt. \qquad (7)$$

To proceed we must obtain dm/dt which is no larger than the Eddington limit given by the balance of radiation pressure and gravity via

$$\eta c^2 dm/dt = 4\pi c G m_P M_{BH}/\sigma \qquad (8)$$

with $\eta$ the radiative efficiency of the disk, c the speed of light, $m_P$ the mass of the proton, and $\sigma$ the Thomson electron cross section. From this balance we obtain the accretion rate as

$$dm/dt = 4\pi G m_P M_{BH}/(\eta\sigma c) \qquad (9)$$

which replaces dm/dt in equation (7) to obtain

$$da/dt = (6)^{0.5} 4\pi G m_P/(\eta\sigma c). \qquad (10)$$

The simplicity of a constant r (i.e. spin independent inner disk radius) as well as a constant $\eta$ (i.e. spin independent efficiency) allows us to integrate this directly to obtain

$$T = \eta\sigma c\, \Delta a/[(6)^{0.5} 4\pi G m_P]. \qquad (11)$$

Given our assumption for the values of spins for the quiescent jet we have

$$\Delta a = 0.1-(-0.1) = 0.2\ , \qquad (12)$$

gives us

$$T = 0.2\eta\sigma c/[(6)^{0.5}4\pi G m_P]. \qquad (13)$$

Using the efficiency of a zero-spin black hole of 0.06 (Bardeen, Press & Teukolsky 1972) and Thomson cross section for the electron, gives us a time of quiescence equal to

$$T = 2.2 \times 10^6 \text{ yrs}. \qquad (14)$$

While most quiescent times reported in Table 1 are compatible with this theoretical result, the one for 3C 219 at 0.5 Myr is too low by a factor of about 4. With one exception, therefore, we conclude that the DDRG sample is accreting near or below the Eddington limit. Observationally, there is a wide range of reported quiescent times for radio galaxies more generally (for a review see O'Dea & Saikia 2021). For large sources the quiescent times are between $10^7$ and $10^8$ years (O'Dea et al 2001; Shulevski et al 2012), which is compatible with sub-Eddington or near-Eddington accretion through the spin transition from counter to co-rotation. For systems like 3C 293, the quiescent time has been estimated to be about $10^5$ years (Joshi et al 2011). This is incompatible with our picture. Of course, we are not excluding the possibility of randomly retriggered activity. But such objects should not show the correlation discussed in this work.

While FRII jets may affect the accretion rate, the specifics of that have been fleshed out in the literature by the first author and others (Singh et al. 2024, sect. 2.2b; Singh & Garofalo 2023 sect. 2.2; Garofalo, Evans & Sambruna 2010 sect. 3.3) so we refer the reader to that literature for more detail. Here we limit ourselves to noting that if the accretion rate drops below the Eddington value, the time to reach a corotating black hole spin value of 0.1 increases. Because no new accretion trigger occurs in the model, the accretion rate cannot increase during the quiescent phase and either persists at the rate at which it was accreting or drops further below that value. Once the 0.1 spin value is reached and a new jet becomes visible, its lifetime is determined by the rate of accretion. If accretion rates are low, the lifetime will be longer for a given supply of fuel. In short, a natural connection exists between the time of quiescence and that of the new jet. This implies that $t_{in}$ and $t_{quies}$ should be correlated as found in Section 2.1. Let us make this statistically robust.

2.2b Origin of the time correlation

Assume two objects once triggered display a certain behavior for different lengths of time. Assume I observe them randomly after they are triggered. Statistically, will more time have passed since the triggering for one of the objects? The answer to this question is that more time will likely have passed when you observe the object whose behavior lasts longer as follows. Imagine that once triggered, object A displays a certain kind of behavior for a time $T_A$ while, once triggered, object B displays the same type of behavior for a time $T_B$. The probability of observing object A and object B while they display that behavior is

$$P_A = T_A/(T_A + T_B), \text{ and } P_B = T_B/(T_A + T_B), \qquad (15)$$

respectively. The average time at which objects A and B will be observed are $T_A/2$ and $T_B/2$. Hence, the expected time at which object A and object B are observed is

$$E_A = P_A T_A/2 \qquad \text{and} \qquad E_B = P_B T_B/2. \qquad (16)$$

The longer the object displays that behavior, the longer the expected time when the object is observed. This applies to radio galaxies in the following way. Assume two radio galaxies once triggered into counter-rotation around a black hole spin down and continuous accretion spins them up into corotation. But during the counter-rotating phase, the accretion rate drops below the Eddington value for one of the radio galaxies but not the other. Say, the accretion rate drops to 10% the Eddington value. As a result of this drop in the accretion rate, the sub-Eddington accreting black hole will reach a prograde black hole spin of 0.1 taking an order of magnitude longer than the Eddington-limited radio galaxy. In other words, its quiescent time will be ten times as long. Since a renewed jet is prescribed to become visible at this spin value, both radio galaxies display jets. However, the radio galaxy that is accreting at its Eddington value will consume its accretion fuel ten times faster than its sub-Eddington counterpart. As a result, the Eddington-limited radio galaxy will display a jet during the corotating phase that is short-lived by an order of magnitude compared to its sub-Eddington counterpart, everything else being equal. Now we imagine observing both radio galaxies when they have jets while in the corotating regime and lobes due to the previous counter-rotating phase. They are, in other words, both DDRGs. Assuming we observe them randomly after the corotating jet is formed, will more time have passed since the triggering of the sub-Eddington radio galaxy? The answer is yes. Assume that radio galaxy B accretes through zero black hole spin at the Eddington rate while radio galaxy A accretes at 10% Eddington. Equations (16) gives us the following expected times since the renewed jet started.

$$E_A = P_A T_A/2 = 0.5T^2_A/(T_A + T_B) = 0.5 \times 100\, T_B^2/(10T_B + T_B) = 50\, T_B^2/(11T_B) = 50T_B/11$$

and

$$E_B = P_B T_B/2 = 0.5T^2_B/(T_A + T_B) = 0.5T^2_B/(10T_B + T_B) = T_B/22.$$

We therefore expect to observe the two DDRG at a time $T_{after}$ after jet forms for radio galaxy B while after a time $100T_{after}$ after the jet forms for radio galaxy A. This argument runs through for radio galaxies whose accretion rates vary by different amounts. If 3 million years is the time it takes for the system in counter-rotation at black hole spin of 0.1 to evolve into a corotating black hole spin of 0.1 (Garofalo et al. 2020), the lowest accretion rate for our sample is about 6% Eddington, i.e 3/52. Of course, $t_{in}$ is not necessarily the dynamical time of the jet and need not be. One simply needs a timescale that scales linearly with actual lifetime of the jet to generate a correlation. We expect the correlation to emerge statistically between the quiescent time and any time associated with the lifetime of the corotating jet. It will be of interest to explore this with future deeper data sets.

In addition to the expected correlation between $t_{in}$ and $t_{qui}$, there is an additional constraint from theory that we use to interpret observations. Beginning in 2020 (Garofalo, Joshi et al 2020), we were able to converge on a deeper understanding of jet morphology in the paradigm; in particular, on the physical reason for the difference between FRII and FRI jets. In the paradigm, FRII jets are triggered by mergers and/or secular processes but not FRI jets, which, instead, are the result of an evolutionary transition. If a black hole is triggered into counter-rotation, it must spin down toward zero spin, and in so doing, the Bardeen-Petterson effect disappears. This means that the black hole will be fed by an accretion disk that may change its orientation by acquiring the orientation determined by the angular momentum of the incoming gas from the greater galaxy. In this context, and if continued accretion occurs, the black hole will spin up in corotation to produce a new jet whose direction will shift compared to the previous jet direction, more directly impacting the interstellar medium and allowing the environment to entrain the jet and turn it into an FRI. However, there is a chance that the counter-rotating black hole may have an angular momentum direction that is not very different from that of the incoming gas from the greater galaxy, which would then not subject the system to a large variation in orientation as the black hole spins down through zero spin. As a result, the new jet in corotation would not shift compared to the counter-rotating phase, and the two jets would share the same axis. But if the jet maintains its orientation, it will not impinge directly on the interstellar medium, and would not be subject to the entrainment that would turn it into an FRI. Hence, there is a physical explanation in the paradigm for why both jet phases should have the same morphology. They should both be FRII jets. In addition to a correlation between time scales, therefore, we also have a prediction that DDRGs should have FRII jet morphology. It is therefore a twofold combination of ideas that support our theoretical picture. While jet re-orientation is key in generating jet entrainment in our paradigm, there are observational differences in magnetic fields around FRII and FRI jets (Bridle & Perley 1984; see Saikia 2022 for a review) that need to be accounted for. There are strictly engine-based features associated with counter-rotation and corotation that produce differences in magnetic field structures. The magnetic fields brought to the black hole in counter-rotation and in corotation differ in that the dragging versus the diffusion of the magnetic field depends on the inner disk edge (Garofalo 2009). An FRII jet in a DDRG, therefore, is predicted to have differences in magnetic field structure compared to FRII radio quasars. We do not explore this further here.

## 3. Discussion and conclusion

We have analyzed the data for DDRG as a way of exploring a new explanation that does not postulate triggering of jets from separate random accretion events, but rather with a picture in which accretion is unabated throughout the various jet phases and in a way that explains jet morphology. Over the past decade, the first author and colleagues have applied the paradigm of counter-rotation to corotation to describe a wide variety of effects of AGN including on star formation, stellar velocity dispersion, and accretion rate, among others. We now discuss the DDRG phenomenon in relation to these other parameters and therefore fit them within the broader context of the paradigm.

DDRG must obviously be powerful enough to create outer lobes that last enough to be visible when the inner jet emerges after quiescence. The most powerful radio galaxies on average in the paradigm are prescribed to emerge in rich environments and progressively less so as one considers group and isolated environments. But the most powerful FRII radio galaxies in the paradigm also enhance star formation and produce a change in accretion from the initial radiatively efficient to advection dominated flow (ADAF - Garofalo, Evans & Sambruna 2010). This slows down the accretion rate compared to the Eddington limit by at least a factor of 100 (i.e. the theoretical boundary between thin-disk accretion and ADAF accretion is .01 the Eddington rate), making the transition time between the two jets on either side of zero black hole spin order 1 billion years. But none of the quiescent times in the sample of Marecki et al. 2021 is that long. The longest quiescent time is roughly 100 Myrs. As a result, none of the objects of Marecki et al. 2021 are prescribed to have rapidly transitioned into ADAF accretion. Hence, the DDRGs analyzed in this work are prescribed in the paradigm to not have outer jets that were the most powerful. This means that on average they should be found in less dense environments like groups. This is a prediction. Because the original FRII jet produced during the counter-rotating accretion phase displayed jet power below some threshold jet power value, its ability to enhance star formation is also prescribed to be limited. The objects of our paper are prescribed, therefore, to not have reached the highest possible values of star formation that are possible in the richest environments. In other words, the model prescribes their positive jet feedback on star formation to be modest. Because DDRGs produce collinear inner and outer jets or lobes, the model prescribes such objects to also be weak in their negative jet feedback on star formation. This is because star formation suppression, in the model, is due to tilted jets that emerge in the corotation phase of accretion (Garofalo, Moravec et al. 2022). This picture then prescribes that our DDRGs should have younger stellar populations compared to the red-and-dead giant elliptical radio galaxies like M87. But jet morphology is also affected by the lack of interaction with the interstellar medium, which explains why the inner, later jet, along with weak feedback features is also of FRII morphology.

While none of our objects have long quiescent times, it is interesting to explore the type of DDRG that may exist in the paradigm in the richest environments. Consider a powerful FRII jet triggered in a rich environment that rapidly transitions its accretion flow from radiatively efficient to ADAF, going from being a high excitation radio galaxy to a low excitation radio galaxy. This would imply an accretion rate that has dropped to at least 1% its Eddington rate, which makes the quiescent period jump to hundreds of millions of years (that is the time to get to a corotating black hole spin value of 0.1). None of the objects in this paper have such features, of course, but it may be possible for DDRGs to have this feature (Mahatma et al. 2019). Because there is a time limit when the lobes of the outer jet are still visible, DDRGs whose FRII jets were the most powerful, are limited by this observational feature and are thus prescribed in the paradigm to be fewer as a result. Another constraint that limits the number of DDRGs in the paradigm is the fact that in the transition through zero black hole spin the accretion flow tends to form a plane whose orientation is random and not constrained by the previous orientation of the accretion disk. This is where we were able to make contact with X-shaped radio galaxies and why they form in isolated environments (Garofalo, Joshi et al. 2020). Most radio galaxies will have jets whose quiescent times are too large to produce either X-shaped radio galaxies or DDRGs. These ideas, therefore, go a long way toward explaining why DDRGs are among the smallest subgroup in terms of numbers in the jetted AGN phenomenon according to our paradigm.

Our picture for DDRGs is not as a matter of principle incompatible with other models for DDRGs such as the "Bow Shock Model" (Brocksopp et al. 2007) which postulates reacceleration in the inner lobes due to the shock produced by the fast-moving inner jet. It is only in tension with models that prescribe other reasons for the triggering of the renewed jet such as changes in the magnetic field needed to launch a jet or disk instabilities (Czerny et al. 2009). In closing we mention triple-double radio galaxies. The model prescribes jet suppression at high enough prograde spin (i.e. $a > 0.7$) if the disk is still thin. If the disk changes to an advection-dominated accretion flow, jet suppression is inhibited and this would ensure a third jet instantiation. This is the only place in the model for two jet quiescent phases in the context of continued accretion. The timescales involved are tens of millions of years, making it difficult for the modeling of triple-double radio galaxies. We do not explore this in any detail here but suggest that such objects may instead be the result of a genuine retriggering of a radio galaxy following some merger event.

## Acknowledgments

We are genuinely indebted to 6 referees for significant contributions in fixing our data, helping us understand the nature of the uncertainties and the statistical significance, and improving the overall sentiment and character of our paper.

## References

Bardeen, J.M., Press, W.H. & Teukolsky, S.A., 1972, ApJ, 178, 347

Bridle, A.H. & Perley, R.A., 1984, ARAA, 22, 319

Brienza, M. et al 2020, A&A, 638, A29

Brocksopp, C., Kaiser, C.R., Schoenmakers, A.P., de Bruyn, A.G., 2007, MNRAS, 382, 1019

Brocksopp, C., Kaiser, C.R., Schoenmakers, A.P., de Bruyn, A.G., 2011, MNRAS, 410, 484

Czerny, B., Siemiginowska, A., Janiuk, A., Nikiel-Wroczynski, B., Stawarz, L., 2009, ApJ, 698, 840

Fabian, A.C., 2012, ARAA, 50, 455

Fanaroff, B.L. & Riley, J.M, 1974, MNRAS, 167, 31

Garofalo, D., Joshi, R., Yang, X., Singh, C.B., North, M., Hopkins, M., 2020, ApJ, 889, 91

Garofalo, D., Moravec, E., Macconi, D., Singh, C.B., 2022, PASP, 134, 114101

Garofalo, D., Evans, D.A., Sambruna, R.M., 2010, MNRAS, 406, 975

Garofalo, D., 2009, ApJ, 699, 400

Jamrozy M., Konar C., Saikia D. J., Stawarz Ł., Mack K.-H., Siemiginowska

A., 2007, MNRAS, 378, 581

Joshi, S.A., Nandi, S., Saikia, D.J., Ishwara-Chandra, C.H., Konar, C., 2011, J.Astrophys.Astr.,32, 487

Konar, C., Hardcastle, M.J., Jamrozy, M., Croston, J. H., Nandi, S., 2012, MNRAS, 424, 1061

Konar, C. & Hardcastle, M.J., 2013, MNRAS, 436, 1595

Konar C., Hardcastle M. J., Jamrozy M., Croston J. H., 2013, MNRAS, 430, 2137

Kukreti, P. et al 2022, A&A, 658, A6

Kuzmicz, A., Jamrozy, M., Koziel-Wierzbowska, D., Wezgowiec, M., 2017, MNRAS, 471, 3806

Machalski J., Chyży K. T., Stawarz Ł. and Kozieł D. 2007 *A&A* 462 43

Machalski, J., Jamrozy, M. & Konar, C., 2010, A&A 510, A84

Machalski J., Jamrozy M., Stawarz Ł., Kozieł-Wierzbowska D., 2011, ApJ,

740, 58

Machalski J., Jamrozy M., Stawarz Ł., Wezgowiec M., 2016, A&A, 595, A46

Mahatma, V.H., et al 2019, A&A, 622, A13

Marecki A., Jamrozy M., Machalski J., 2016, MNRAS, 463, 338

Marecki, A., Jamrozy, M., Machalski, J., Pajdosz-Smierciak, U., 2021, MNRAS, 501, 853

Moderski, R., Sikora, M., Lasota, J-P., 1998, MNRAS, 301, 142

O'Dea, C.P. & Saikia, D.J., 2021, The Astronomy & Astrophysics Review, 29, 3

O'Dea, C.P., Koekemoer, A.M., Baum, S., A., Sparks, W.B., Martel, A.R., Allen, M.G., Macchetto, F.D., Miley, G.K., 2001, AJ, 121, 1915

Orru', E. et al 2015, A&A, 584, A112

Saikia, D.J., 2022, J.Astrophys.Astr., 43, 2

Saikia, D.J. & Jamrozy, M., 2009, BASI, 37, 63

Schoenmakers, A.P., de Bruyn, A.G., Rottgering, H.J.A, van der Laan, H., Kaiser, C.R., 2000, MNRAS, 315, 371

Shulevski, A., Morganti, R., Oosterloo, T., Struve, C., 2012, A&A, 545, A91

Siemiginowska , A. et al 2012, ApJ, 750, 124

Sikora, M., Stawarz, L., Lasota, J-P, 2007, ApJ, 658, 815

Singh, C.B., Williams, M., Garofalo, D., Rojas-Castillo, L., Taylor, L., Harmon, E., 2024, Universe, 10, 319

Singh, C.B. & Garofalo, D., 2023, JHEAP, 39, 21

Sudheesh, T.P. et al 2025, JAstrophysAstron, 46, 74

Wilson, A.S. & Colbert, E.J.M., 1995, ApJ, 438, 62

Wolnik, K., Jurusik, W. and Jamrozy, M., 2024, A&A, 691, A76